\documentclass[submission,copyright,creativecommons,noderivs,noncommercial]{eptcs}
\providecommand{\event}{Fourth Workshop on Formal Methods\\ for Autonomous Systems}
\usepackage[T1]{fontenc}        % Recommended with pdflatex
\usepackage{amssymb}
\usepackage{amsmath}
\usepackage{todonotes}
\usepackage{csquotes}
\usepackage[per-mode=symbol, binary-units=true]{siunitx}
\usepackage{caption}
\usepackage{subcaption}
\usepackage{wrapfig}
\usepackage{makecell}
\usepackage{fdsymbol}
\usepackage{float}
\edef\restoreparindent{\parindent=\the\parindent\relax}
\newcommand{\doi}[1]{\textsc{doi}: \href{http://dx.doi.org/#1}{\nolinkurl{#1}}}
\restoreparindent
\usetikzlibrary{arrows}
\usetikzlibrary{trees,shapes,decorations,fit}
\usetikzlibrary{shapes.multipart,arrows.meta}
\tikzset{%
	>={Latex[width=2mm,length=2mm]},
	base/.style = {rectangle, rounded corners, draw=black,
		minimum width=4cm, minimum height=1.5cm,
		text centered, font=\footnotesize},
	activityStarts/.style = {base, fill=blue!30},
	inner/.style = {base, dashed, line width=1.5pt},
	activityRuns/.style = {base, draw=green!70!black, line width=1.5pt},
	process/.style = {base, minimum width=2.5cm, line width=1.5pt},
}

\DeclareMathAlphabet{\mathpzc}{OT1}{pzc}{m}{it}



\graphicspath{{figures}}

\usepackage{adjustbox}

\usepackage{array}
\newcolumntype{P}[1]{>{\centering\arraybackslash}p{#1}}
\newcolumntype{M}[1]{>{\centering\arraybackslash}m{#1}}

\usepackage[usestackEOL]{stackengine}
\usepackage[inline]{enumitem}

\title{Towards Runtime Monitoring of Complex System Requirements for Autonomous Driving Functions\thanks{\small{This research was funded by the German Federal Ministry of Economic Affairs and Climate Action (BMWK) through the "KI Wissen" project under grant agreement No. 19A20020M.	}}}
\author{Dominik Grundt
\email{dominik.grundt@dlr.de}
\and
Anna Köhne
\email{anna.koehne@dlr.de}
\and
Ishan Saxena
\email{ishan.saxena@dlr.de}
\and
Ralf Stemmer
\email{ralf.stemmer@dlr.de}
\and
Bernd Westphal
\email{bernd.westphal@dlr.de}
\and
Eike Möhlmann
\email{eike.moehlmann@dlr.de}
}

\def\titlerunning{Towards Runtime Monitoring of Complex System Requirements for Autonomous Driving Functions}
\def\authorrunning{D. Grundt, A. Köhne, I. Saxena, R. Stemmer, B. Westphal \& E. Möhlmann}
\def\copyrightholders{D. Grundt, A. Köhne, I. Saxena,\\ R. Stemmer, B. Westphal \& E. Möhlmann}

\begin{document}
\maketitle
\begin{center}
	{\small German Aerospace Center \\ 
	Institute of Systems Engineering for Future Mobility \\
	Oldenburg, Germany}
\end{center}

\vspace*{.5cm}
\begin{abstract}
 Autonomous driving functions (ADFs) in public traffic have to comply with complex system requirements that are based on knowledge of experts from different disciplines, e.g., lawyers, safety experts, psychologists. 
 In this paper, we present a research preview regarding the validation of ADFs with respect to such requirements.
 We investigate the suitability of Traffic Sequence Charts (TSCs) for the formalization of such requirements and present a concept for monitoring system compliance during validation runs. 
 We find TSCs, with their intuitive visual syntax over symbols from the traffic domain, to be a promising choice for the collaborative formalization of such requirements.
 For an example TSC, we describe the construction of a runtime monitor according to our novel concept that exploits the separation of spatial and temporal aspects in TSCs, and successfully apply the monitor on exemplary runs.
 The monitor continuously provides verdicts at runtime, which is particularly beneficial in ADF validation, where validation runs are expensive.
 The next open research questions concern the generalization of our monitor construction, the identification of the limits of TSC monitorability, and the investigation of the monitor's performance in practical applications. 
 Perspectively, TSC runtime monitoring could provide a useful technique in other emerging application areas such as AI training, safeguarding ADFs during operation, and gathering meaningful traffic data in the field.
 
\end{abstract}

\section{Introduction}
\label{sec:intro}
	
The importance of autonomous driving functions (ADFs) for future mobility is increasing rapidly~\cite{Ma2020,Yurtsever2020}.
ADFs have to safely operate in complex and dynamic traffic environments while their internal behavior is often not sufficiently known, e.g.,\ due to the presence of AI-based components. Hence, system validation during their development is necessary for homologation. Here, validation consists of evaluating a representative set of system runs for compliance with system requirements. 
A promising approach for ADF validation in the automotive domain is scenario-based testing~\cite{es3,neurohr2020,pegasus}.
This method is used for validation with respect to technical system requirements (e.g.,\ limits on maximum acceleration). However, it cannot yet be used for more complex system requirements. 
In order to act safely in traffic environments, ADFs also need to take into account complex world knowledge about traffic rules and social norms regarding the interaction of traffic participants, resulting in more complex system requirements. Knowledge of several stakeholders, e.g.,\ lawyers, safety experts and psychologists, needs to be integrated into ADF system requirements. In many cases, these system requirements concern the temporal evolution of spatial relations between traffic participants and their environment. In order to objectify requirements derived from such knowledge, formalization is important. 
Thus, for ADF validation, there is a need for a formalization language that can usefully formalize knowledge from several stakeholders to specify spatio-temporal system requirements and a suitable method that checks the system's behavior against these requirements.
Since the validation of ADFs is a time and resource intensive process, runtime monitoring would be particularly beneficial due to continuous calculation of verdicts. Specifically, the continuously calculated verdicts may provide additional insights into cause and effect of violations and can be used as early termination criteria for test runs.

In this paper, we present a research preview for formalizing spatio-temporal system requirements from the automotive domain and a corresponding runtime monitoring concept for the validation of ADFs.
To this end, we investigate whether the specification language Traffic Sequence Charts (TSCs)~\cite{atr117,Damm2018} can be used for formalizing such automotive system requirements and how to perform runtime monitoring for this formalism.
The paper is organized as follows: In \autoref{sec:relatedwork}, we discuss related work regarding specification languages and runtime monitoring. In \autoref{sec:tsc}, we investigate the suitability of the TSC language for our purpose. In \autoref{sec:rmconcept}, we present and discuss our concept for runtime monitoring of TSCs. Finally, in \autoref{sec:conclusion}, we conclude and discuss future work.

\section{Related Work}
\label{sec:relatedwork}
In this section, we discuss literature regarding specification languages for traffic scenarios and runtime monitoring. 
\textit{Specification Languages.} In recent years there has been a trend towards the visual specification of automotive system requirements. Visual specification languages enable a more intuitive understanding and more concise representation of system requirements as compared to textual or symbolic languages. They allow experts from different domains to define, discuss, adapt and further derive system requirements or scenarios, without a deep understanding of symbolic logic. The MLSL formalism \cite{Schwammberger2021} enables visual specification of requirements for system controllers of cooperative systems for urban road intersections with a focus on reasoning about authority of movement. The specification language M-SDL \cite{msdl19} focuses on sampling a large number of concrete test cases based on abstract scenarios. While it allows visualization of scenarios, it lacks semantics for its visual elements. The same also applies to the SCENIC framework \cite{Fremont_2019}, which can generate large numbers of test scenarios based on scenario descriptions that can also be displayed visually. Here, a scene is generated from code description. In contrast, the TSC language \cite{atr117,Damm2018} does not have a specific focus on a particular type of traffic scenarios and has been applied to the specification of test scenarios \cite{es3,pegasus} as well as system requirements \cite{Becker2020}.
While the visual specification of scenarios and system requirements is already used in research and industry, there still is a need for the development of runtime monitoring for visually specified system requirements.

\textit{Runtime Monitoring} is an established technique for validating system behavior with respect to system requirements. Runtime monitoring methods for temporal logics such as Linear Temporal Logic (LTL) and Signal Temporal Logic (STL) are available in literature. In \cite{bauer2011}, the authors present such a method for LTL. In the last decade, there has been an increasing interest in runtime monitoring of STL as presented in the works \cite{Bartocci2018,Deshmukh2015RobustOM,Donze13,donze2010robust,jakvsic2015signal}. Additionally, extensions to STL and its runtime monitoring have been developed for spatial as well as spatio-temporal requirements \cite{Haghighi2015,stsl2021,nenzi2017}. In \cite{ferrere2019}, timed automata have been used for the verification of real-time specifications based on the logic MITL. In \cite{rupak2021}, based on programmed scenarios and available temporal properties, monitors are programmed manually and used to analyze streams generated by the system. Closely related to our concept is the recent work \cite{Zapridou2020}, where runtime monitoring for an Adaptive Cruise Control of an autonomous vehicle was presented. Their goal was to monitor robustness of PID controller requirements (invariant over speed and distance) inside the vehicle. A specification was created using STL and the temporal behavior of the controller was monitored. Despite the presence of literature for runtime monitoring of spatio-temporal system requirements, there is a gap between visual specification and existing methods. In this paper, we close this gap by presenting a concept for runtime monitoring of visualized complex spatio-temporal system requirements.

%\vspace{-.2cm}
%\pagebreak
\section{TSC for Formalization of Complex System Requirements}
\label{sec:tsc}

As discussed above, there is a need for a specification language that is able to express complex system requirements for autonomous driving functions (ADFs) intuitively, s.t. they can be interpreted and discussed by stakeholders from a wide variety of fields. 
As is inherent to the traffic domain, system requirements for ADFs (e.g., the execution of a specific maneuver or the compliance with traffic rules) often concern the temporal evolution of spatial relations (like relative positions and distances) between, and the state (like the velocity or activation of a turn signal) of, traffic participants and their environment.

The Traffic Sequence Chart (TSC) language is a formalism that provides visualization of spatio-temporal system requirements with a rigorous formal semantics and tool support~\cite{becker2022} for automatic translation to a machine-interpretable format. 
In this section, we investigate the suitability of the TSC language for the formalization of complex spatio-temporal system requirements for ADF validation and give a simplified overview of syntax and semantics of TSCs. For a detailed account of the TSC language, refer to \cite{atr117}. Regulations for ADFs are based on expert knowledge from a variety of domains.
For discussions between stakeholders, system requirements are often only specified informally in natural language~\cite{un2021}. We now provide an example of such an informal specification. For this example, the system we want to validate is an ADF controlling a car. We formulate a requirement for runs of the system, which gets activated when the car approaches an obstacle in the right lane of a two-lane road with no other traffic: 

\emph{``The car should change to the left lane, pass by the obstacle and return to the right lane in an orderly fashion within less than 45 seconds, while always keeping a safe distance to the obstacle."}

\begin{figure}[htb]
	\centering
	\includegraphics[width=0.99\textwidth]{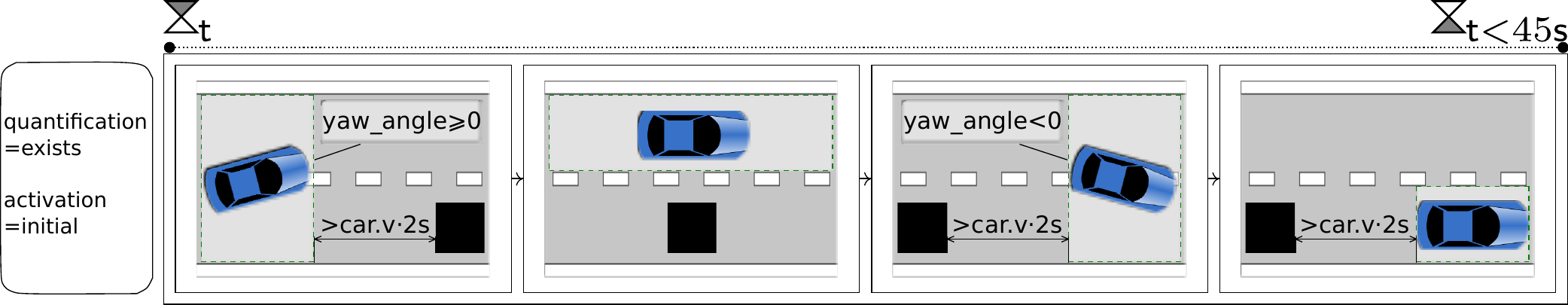}
	\caption{Example TSC of a pass-by maneuver.}
	\label{fig:TSC}
\end{figure}

To enable objective verdicts, such informal descriptions need to be formalized. We demonstrate a formalization as a TSC in~\autoref{fig:TSC}.

As in \autoref{fig:TSC}, every TSC consists of a Header and a Chart. In the Header, the start and end of the considered time interval are quantified as the so called \emph{activation mode} and \emph{time quantification}. In our example, the activation mode is \emph{initial} and the time quantification is \emph{exists}, indicating that there must be some initial (starting at activation) time interval in which the car behaves as specified in the Chart. 
Charts are constructed from Invariant Nodes, each inscribed with a Spatial View, and may be annotated with an Hourglass Constraint, specifying requirements regarding the duration of the corresponding time interval.
In our example, the Chart is a \emph{sequence} of four Invariant Nodes with an Hourglass Constraint of $<45s$.
Within a Spatial View, objects (like cars, obstacles and roads) are represented using object symbols. Constraints on the spatial relations between the represented objects are indicated by the relative positions of their symbols as well as \emph{distance lines} explicitly constraining the allowed distance between objects. Furthermore, objects have attributes (like velocity or yaw angle), which may also appear in constraints. 

In the Spatial Views of our example, a car is represented by a blue car symbol and a stationary obstacle by a black square. The first Spatial View is satisfied when the car is at a safe distance of more than $\mathit{car}.v\cdot 2\text{s}$ (distance covered in 2 seconds) behind the obstacle and is oriented towards the left side of the road, and the second when the car is on the left lane. The third Spatial View is satisfied when the car is at least the specified safe distance ahead of the obstacle and is oriented towards the right side of the road, and the fourth when the car is on the right lane and still at least the specified safe distance ahead of the obstacle.
The TSC is satisfied for a given run of a system (starting at the activation of our TSC) if there is an end time $t$ less than 45 seconds, and times $0=t_0<t_1<t_2<t_3<t_4=t$ such that for $i\in\{1,2,3,4\}$, on $[t_{i-1},t_i)$, the $i$-th Spatial View is satisfied.

Our example shows that TSCs can, in principle, be used for formalizing complex spatio-temporal system requirements. Additionally, it demonstrates the intuitive visual syntax of TSCs, which satisfies the need for multi-stakeholder collaboration. Besides sequentially connecting Invariant Nodes (or more generally Charts), the formalism provides the syntax and semantics to express \emph{Negation}, \emph{Concurrency} and \emph{Choice} (conceptually logical not, and, or) as well as implications between Charts. In particular, the expression of implications together with a universal quantification of the considered time interval in the Header is useful for the formalization of rules like ``whenever a car has not indicated a lane change recently, it must stay on its current lane".

In conclusion, the TSC language is a promising choice for formalizing complex spatio-temporal system requirements in the context of validating ADFs.

\section{Runtime Monitoring for the TSC language}
\label{sec:rmconcept}

In this section, we define our notion of runtime monitoring and based on that describe a novel concept for runtime monitoring of visually specified complex spatio-temporal automotive system requirements, using the TSC language.
We define runtime monitoring as the task of checking the satisfaction of a correctness property given a run of a system. In particular, a runtime monitor must decide whether the run satisfies or violates the correctness property as early as possible. At runtime, the monitor iteratively evaluates the current initial segment (prefix) of the run for the following properties: a prefix is called \emph{satisfying} if every continuation satisfies the property, \emph{violating} if no continuation satisfies the property or \emph{inconclusive} otherwise. It returns \textit{violating} or \textit{satisfying} as verdict as soon as either is detected. In TSC runtime monitoring, the correctness property is given by satisfaction of a TSC. %\remove{by a system run}. %For simplicity we assume from now on, that there is only one possible valuation of global id variables. Otherwise the method described below may and would have to be applied to different valuations separately.
\newpage
\begin{figure}[htb]
	\centering
	\resizebox{0.99\textwidth}{!}{\input{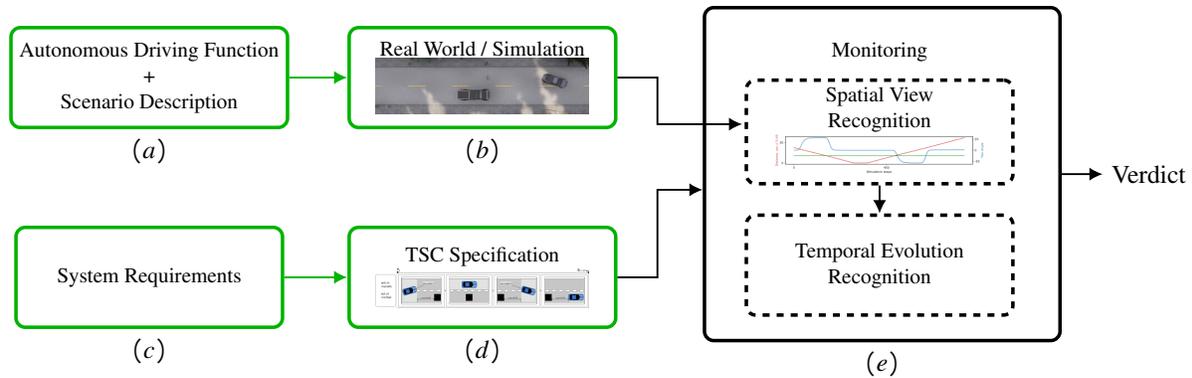}}
	\caption{ TSC runtime monitoring concept}
	\label{fig:concept}
\end{figure}
In \autoref{fig:concept}, we present an overview of our concept for the runtime monitoring for TSCs. The green boxes~$(a\text{-}d)$ and arrows describe artifacts and methods that we consider as given. The contents of $(a)$ an autonomous driving function~(ADF) and scenario description, $(b)$ corresponding runs in a simulation or real world environment, and $(c)$ system requirements are already present in the scenario-based testing approach. Based on our discussion in~\autoref{sec:tsc}, we propose the use of TSCs for the formalization of system requirements~$(d)$. The focus of this section is on the dashed black boxes~$(e)$. 
The TSC language specifies spatial and temporal aspects explicitly on different levels through Spatial Views and Chart structure respectively. Our concept exploits this property by analyzing them separately through Spatial View Recognition (SVR) and Temporal Evolution Recognition (TER), at runtime. Together, SVR and TER build the runtime monitor. Following, we describe SVR and TER using our example TSC~(\autoref{fig:TSC}) and discuss upcoming challenges for the successful realization of our concept.

We define SVR as a method to identify which Spatial Views are satisfied at any specific point of time. SVR is provided, as input, with sensor data based attributes (e.g.,\ a car's yaw angle) extracted from real world/simulation sensors. Additionally, SVR has access to a collection of spatio-temporal system requirements specified using TSCs along with a unique ID for every occurring Spatial View. It serves as a connection between formal specifications and physical/virtual sensor information. The SVR continually produces as output the set of IDs of the Spatial Views that are currently satisfied. Here it should be noted that the SVR output at a given time can include Spatial Views from multiple TSCs or may even be empty, if no Spatial View is satisfied. Additionally, each output is annotated with a timestamp.

If the required attribute values are not provided directly, then the SVR must calculate them internally from the available input data. For our example TSC, the longitudinal distance between car and obstacle as well as the location of the car on the road and with respect to the obstacle will need to be calculated. Spatial Views encode conjunctions of arithmetic constraints on real valued attributes. Once the calculation of attributes is finished, SVR identifies the set of Spatial Views that are currently satisfied and passes this information to TER. 

TER iteratively receives the output of SVR as input. Based on this analysis of the spatial aspect of the system requirements, TER analyzes the temporal aspect and decides whether the system run, up to the current time, is \emph{satisfying}, \emph{violating} or \emph{inconclusive} and produces the appropriate output.
For our example TSC, we have done this using a timed automaton. Its construction is based on keeping track of which initial segment(s) of the TSC, the system run up to the current time might correspond to in the following sense:
A (prefix of a) system run of duration $d$ \emph{fits the TSC up to} $j$ for $j\in\{0,1,2,3,4\}$, iff there exist times $0=t_0<t_1<\ldots<t_{j-1}\leq d <t_j$ such that for $i\in\{1,2,\ldots,j\}$ on $[t_{i-1}, t_{i})$, the $i$-th Spatial View is satisfied. By definition, a run satisfies the TSC exactly if there is a prefix of this run of duration less than $45$, which fits the TSC up to $4$.
It can be shown that a prefix of a run is: \emph{satisfying}, exactly if the prefix already satisfies the TSC; \emph{inconclusive}, if it is not satisfying, has a duration less than $45$ and fits the TSC up to $j$ for some $j<4$, and \emph{violating} otherwise. 

The runtime monitor producing the output \emph{satisfying}, for a specific run, implies that the ADF successfully performs the specified maneuver. In case the output produced is \emph{violating}, this implies that the ADF violates the system requirements. Additionally, the time at which the output \emph{violating} is produced identifies the point of time during the run at which the requirement is first violated. Hence, connecting the runtime monitor to a source of relevant data allows first statements to be made about the validity of the ADF.

\begin{figure}[htb]
	\centering
	\includegraphics[width=0.99\textwidth]{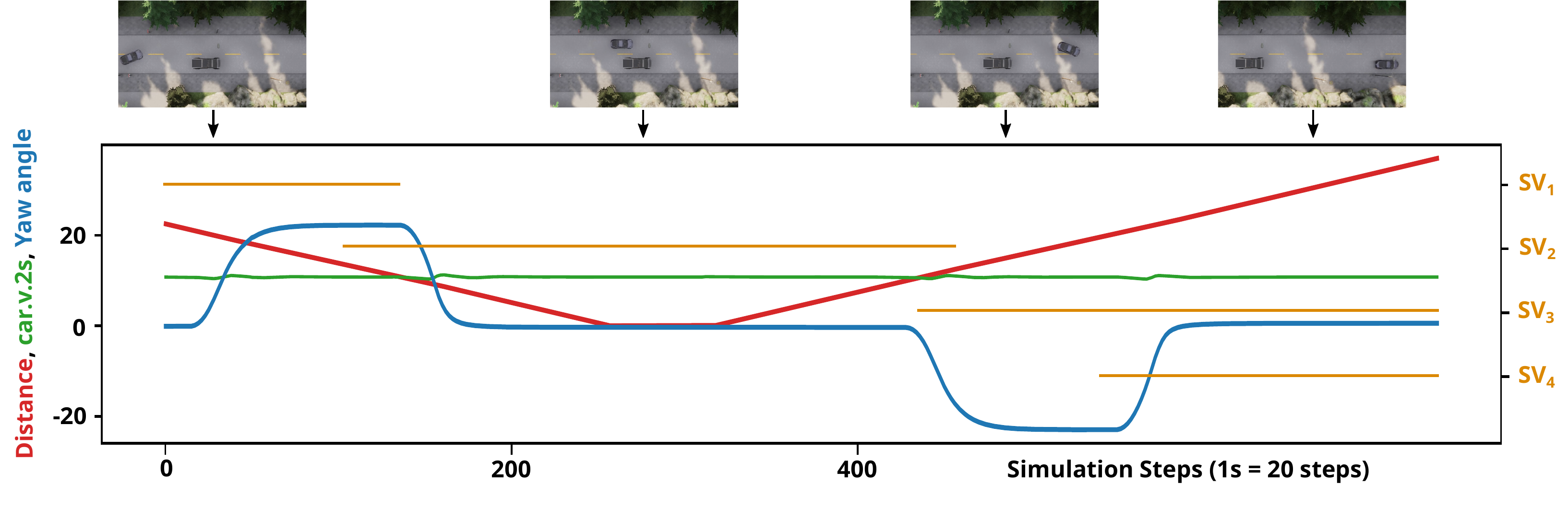}
	\caption{TSC-satisfying simulation run} 
	\label{fig:successfull_maneuver}
\end{figure}

To test the feasibility of our concept, we carried out exemplary runs in the autonomous driving simulator CARLA~\cite{Dosovitskiy17}. The runs start with a car approaching an obstacle such that the example requirement as described in~\autoref{sec:tsc} is activated. We continuously computed relevant attribute values from available simulation data and visualized their evolution for one such run in~\autoref{fig:successfull_maneuver}. The red graph represents the distance between car and the stationary obstacle over time, while the blue graph represents variation of the car's yaw angle over time. The green graph describes, at each point of time, the safe distance of the car to the obstacle as defined in the system requirement. Above these graphs, we depict images from the simulation for selected points of time to illustrate the simulation run. 
The run will be compared against the Spatial Views of the TSC shown in \autoref{fig:TSC}. Their semantics remain the same as described in~\autoref{sec:tsc}. 
The orange lines specify the times during the run at which a particular Spatial View is satisfied. This example shows the possibility to map input data to Spatial Views. Using the output from SVR, TER is able to evaluate the temporal aspect. Hence, SVR and TER together constitute a monitor for complex automotive system requirements specified as TSCs. 

\textit{Discussion.}
We now discuss current research challenges that have been identified for the successful implementation of our concept. Execution of SVR on resource-constrained hardware at runtime is currently limited by the demand for time and resource efficient calculation of relevant attribute values and performing satisfaction checks for a potentially large set of individual Spatial Views. 

Furthermore, to ensure (efficient) monitorability  of requirements, there is a need to consider the available sensor data when choosing specific attributes for the specification.  
\newpage
The construction of TERs for general Chart structures and Headers is also a current topic of research. We have constructed TERs for some Charts using timed automata and are currently considering the use of Petri Nets in a more general construction. It needs to be investigated for which subsets of the TSC language these approaches are feasible and further, which subset of the TSC language is monitorable in general.  
For this purpose, Chart structures using the operations choice, concurrency, negation and implication will need to be considered. 

Overall, SVR and TER must be efficiently implemented to be useful on embedded hardware. 
The efficiency of the monitor can be further improved by additional communication between SVR and TER, e.g., by providing feedback to SVR about which specific Spatial Views are relevant for future verdicts.
\section{Conclusion \& Future Work}
\label{sec:conclusion}
In this paper, we investigated and consequently demonstrated the suitability of the Traffic Sequence Chart~(TSC) language for formalizing complex automotive system requirements obtained from multiple stakeholders in the context of autonomous driving function~(ADF) validation. We also presented a concept towards runtime monitoring of complex spatio-temporal system requirements, visually specified as TSCs and tested its feasibility. Our novel concept takes advantage of the TSC structure by separating spatial and temporal aspects of complex system requirements to support system validation. For TSC runtime monitoring, we introduce the monitor components, Spatial View Recognition~(SVR) and Temporal Evolution Recognition~(TER), along with their interfaces. Our concept bridges a gap between visual specifications and runtime monitoring of complex automotive system requirements and combines their advantages.
The concept provides a first step towards application of TSC monitoring for validation of ADFs. The availability of runtime monitors for TSCs could significantly increase automation and efficiency of scenario-based testing methods, and thereby support the homologation of ADFs. The concept can also be useful for plausibilization, safeguarding, and training of AI-based driving functions.

For a future application of our concept to scenario-based testing of ADFs, the challenges discussed in \autoref{sec:rmconcept} need to be addressed.  Also, integrated monitoring of TSCs that characterize proper driving or critical traffic scenarios could be used to auto-trigger switching of automated driving levels, provide warning to the driver or to the driving function developers. Additionally, our concept can also help to improve the efficiency of AI driving function training, in particular for Active Learning and Reinforcement Learning~\cite{mallozzi2019}, e.g., by supplying an additional criterion for early termination of training runs. Last but not least, the above application cases are not limited to only the automotive domain. TSCs may also be used for the formalization of requirements from other traffic domains such as maritime, railway and aerospace. While some of these may benefit from a domain-specific extension of the Spatial View formalism, our monitoring approach is, in principle, independent of the domain.

\newpage
\nocite{*}
\bibliographystyle{eptcs}
\bibliography{lib}
\end{document}